\documentclass[aps,twocolumn,amsmath,amssymb,nobibnotes,showkeys,showpacs,prb]{revtex4-1}

\usepackage{dcolumn}
\usepackage{bm}
\usepackage[T1]{fontenc}
\usepackage{amsmath}
\usepackage{graphicx}
\usepackage{tabularx}
\usepackage{soul}
\usepackage[normalem]{ulem}

\usepackage{hyperref}
\hypersetup{colorlinks=true, linkcolor=blue, citecolor=blue, urlcolor=blue, pdftitle={Spin-texture induced by oxygen vacancies is strontium perovskites surfaces: A theoretical comparison between SrTiO$_3$ and SrHfO$_3$ (001)}, pdfauthor={A. C. Garcia-Castro}}

\begin{document}
\title{Spin-texture induced by oxygen vacancies in Strontium perovskites (001) surfaces: A theoretical comparison between SrTiO$_3$ and SrHfO$_3$}
\author{A. C. Garcia-Castro$^{1}$}
\email{a.c.garcia.castro@gmail.com}
\author{M. G. Vergniory$^{2}$}
\author{E. Bousquet$^{1}$}
\author{A. H. Romero$^{3}$}
\email{alromero@mail.wvu.edu}
\affiliation{$^1$Physique Th\'eorique des Mat\'eriaux, Universit\'e de Li\`ege, B-4000 Sart-Tilman, Belgium}
\affiliation{$^2$Donostia International Physics Center, 20018  Donostia/San Sebastian, Spain}
\affiliation{$^3$Physics Department, West Virginia University, WV-26506-6315, Morgantown, USA}

\begin{abstract}
The electronic structure of SrTiO$_3$ and SrHfO$_3$ (001) surfaces with oxygen vacancies is studied by means of first-principles calculations. 
We reveal how oxygen vacancies within the first atomic layer of the SrTiO$_3$ surface (i) induce a large antiferrodistortive motion of the oxygen octahedra at the surface, (ii) drive localized magnetic moments on the Ti--3$d$ orbitals close to the vacancies and (iii) form a two-dimensional electron gas localized within the first layers.
The analysis of the spin-texture of this system exhibits a splitting of the energy bands according to the Zeeman interaction, lowering of the Ti--3\emph{d$_{xy}$} level in comparison with \emph{d$_{xz}$} and \emph{d$_{yz}$}  
and also an in-plane precession of the spins.
No Rashba-like splitting for the ground state neither for ab--initio molecular dynamics trajectory at 400K is recognized as suggested recently by A. F. Santander-Syro \emph{et al.} \cite{Santander-Syro2014}.
Instead, a sizeable Rashba-like splitting is observed when the Ti atom is replaced by a heavier Hf atom with a much larger spin-orbit interaction. However, we observe the disappearance of the magnetism and the surface two-dimensional electron gas when full structural optimization of the SrHfO$_3$ surface is performed. Our results uncover the sensitive interplay of spin-orbit coupling, atomic relaxations and magnetism when tuning these Sr-based perovskites.
\end{abstract}


\maketitle

\section{Introduction}
Oxide perovskites exhibit a very large variety of physical properties, which allow their use in a wide range of applications such as spintronics or thermoelectricity \cite{phys-oxides2004, 4160091, bibes2014}. 
In particular, during the last decade it has been observed new and exotic functional properties at the surface of oxide crystals with no parallel to their bulk counterpart \cite{Zubko2011, Martin201089}.
In this field of multifunctional oxide materials, SrTiO$_3$ (STO) has attracted a lot of attention due to its wide range of physical and chemical properties that goes from superconductivity \cite{Reyren31082007} to engineered ferroelectricity \cite{Haeni2004}. 
In various studies a two-dimensional electron gas (2DEG) has been reported at the (001) surface of STO crystal, 
both theoretically\cite{PhysRevB.86.195119} and experimentally by using angle-resolved photoemission spectroscopy (ARPES)\cite{Santander-Syro2011,STO-quasiparticle}, which appeared to be induced by the presence of oxygen vacancies and explained by an electronic charge redistribution at the surface. 
The dynamics of the Ti--$d$ electron quasiparticle on this surface have been experimentally characterized and the appearance of cooperative effects of the electrons has been reported, giving rise to an orbital ordering itself leading to a Rashba-like splitting at the $d_{xy}$ - $d_{xz/yz}$ bands crossing close to the Fermi level\cite{STO-quasiparticle}. Recently, a mixed character of the confined electrons of the (001) surface of STO displaying a complex spin texture at the Fermi surface has been reported.
This unexpected result was ascribed to a Zeeman interaction plus Rashba-like splitting of the lowest \emph{d$_{xy}$} owed as well to oxygen vacancies\cite{Santander-Syro2014}. 

Similar Rashba-effect\cite{rashba1984} in oxides has been observed in the LaAlO$_3$/SrTiO$_3$ interface \cite{PhysRevLett.104.126803, PhysRevLett.109.196401, PhysRevB.87.161102, Hurand2015} or in SrTiO$_3$-  and KTaO$_3$-based transistors \cite{PhysRevLett.108.206601, PhysRevB.80.121308}. 
For these single perovskite-based devices, a $k$-cubic dependence of the splitting was found in contrast to the linear+cubic dependence of the Rashba splitting at the LaAlO$_3$/SrTiO$_3$ interface due to the $d_{xy}$ -- $d_{xz/yz}$ multiorbital nature of the lowest bands \cite{PhysRevB.87.161102}. 
As in the case of Au(111) surface, where it has been shown that even when the surface states exhibit a mainly $p$--orbital character, the $d$--orbitals drives the Rashba splitting and dictates the direction of the spin by means of the orbital-angular momentum (OAM) \cite{PhysRevB.86.045437, Park20156}. 
Therefore, the $d$-orbitals of the transition metal atom of perovskite oxides can be the key parameter for a large and controllable electron band splitting in future spintronic devices \cite{PhysRevB.86.045437,Shanavas2014}. 
This later property gives an additional degree of freedom to the rich field of multifunctional perovskites.
However, the recent findings of a giant spin-splitting at the STO surface, opens some questions about the nature of the spin-splitting in this type of materials, such as: 
What is the origin of this giant-splitting? 
Could it be manipulated or observed in similar compounds? 
What is the role of the spin-orbit coupling strength in the Rashba splitting of $d$-orbitals in oxide perovskites? 
Since magnetism at the surface was reported, how does it couple with the splitting?.
From the theoretical point of view, it is well known that density-functional theory (DFT) formalism provides invaluable information related to the structural, magnetic and electronic structure of oxide  interfaces, superlattices and surfaces \cite{MRS:9125030,Nick-2014,Zubko2011,PhysRevB.87.161102}.
This has been particularly true for the SrTiO$_3$/LaAlO$_3$ interface \cite{Nick-2014, PhysRevX.3.021010} or the STO surface \cite{PhysRevX.3.021010} where DFT helped in understanding the origin of the 2DEG through the polar catastrophe scenario.
In the present  paper we address the previous questions by performing such DFT calculations on the (001) STO surface with oxygen vacancies in order to (i) shed light onto this novel experimental finding and (ii) to scrutinize the microscopic origin of the giant Rashba spin-splitting and the Ti $d$-orbitals spin-texture.



 \begin{figure}[htb!]
 \centering
 \includegraphics[width=7.5cm,keepaspectratio=true]{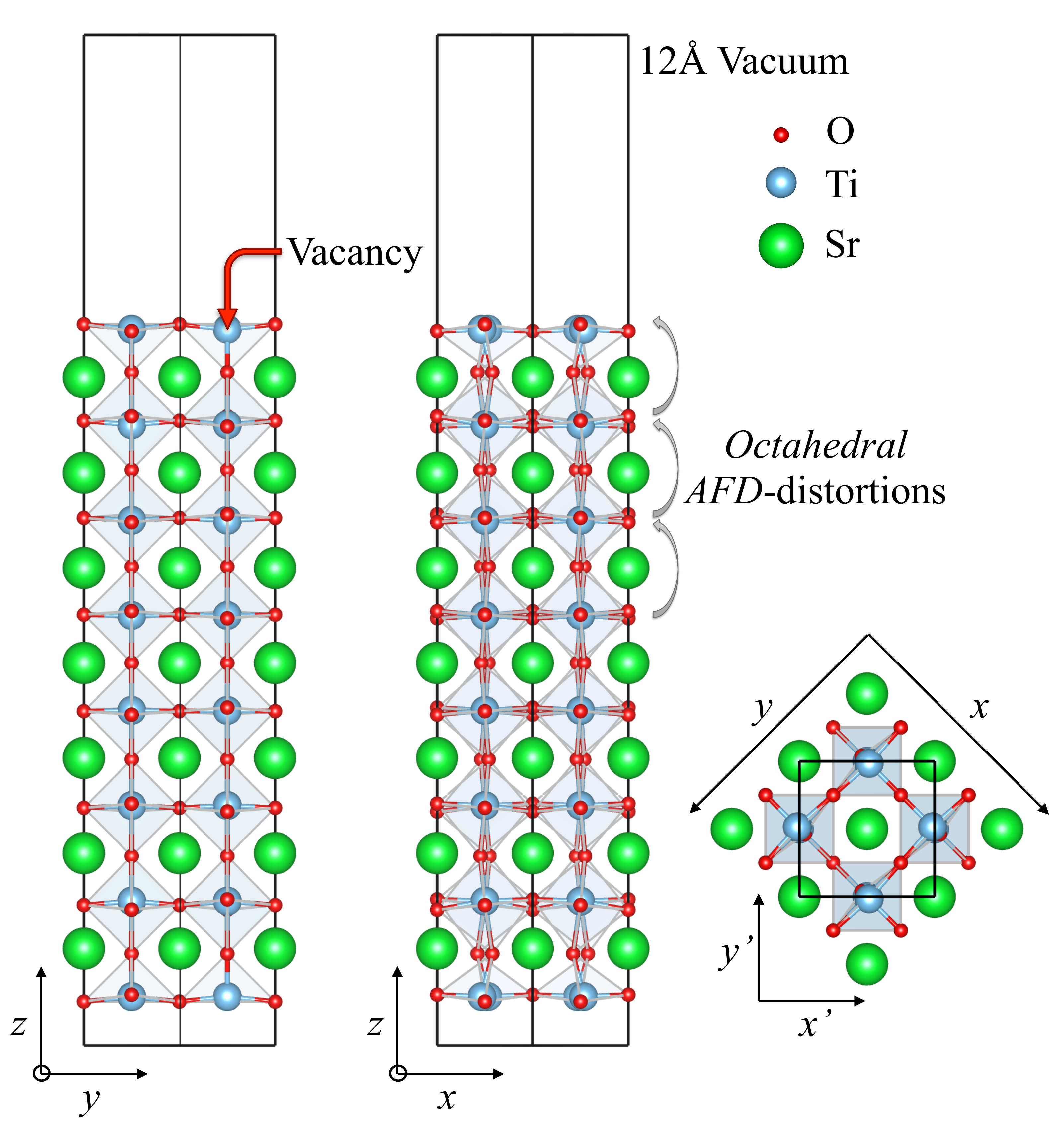}
 \caption{(Color online) Schematic view of the SrTiO$_3$: TiO$_{2-\delta}$ (001) surface used in the calculations with an oxygen vacancy before (left) and after (right) atomic relaxations.
 The amplification of the AFD distortions due to the oxygen vacancy can be clearly appreciated. 
 Additionally, a surface induced off-centering of the Ti along the $z$ direction is observed with respect to the oxygens (rumpling), inducing an electric dipole. 
 In the right bottom part we present a top view of the unit cell in the $xy$-plane.
 This figure was elaborated with the support of the VESTA code \cite{VESTA}. }
 \label{fig:STO-d}
\end{figure}

\section{Computational Details}
The first-principles calculations have been performed within the DFT as implemented in the Vienna Ab-initio Simulation Package (VASP) \cite{Kresse1996,Kresse1999}.  
The interaction between ion cores and valence electrons was treated by the projector augmented wave method\cite{Blochl1994}.
The orbitals considered as valence electrons in the atomic pseudo-potentials were: Sr (4$s^2$4$p^6$5$s^2$),  Ti (3$s^2$3$p^6$4$s^2$3$d^2$), Hf (5$s^2$5$p^6$5$d^2$6$s^2$), and O (2$s^2$2$p^4$). 
To describe the exchange-correlation energy, we used the General Gradient Approximation (GGA) with the Perdew-Burke-Ernkzerhof for Solids (PBEsol) parameterization \cite{Perdew2008}. 
In order to account for the magnetic character of the system we performed spin polarized calculations. 
The Hamiltonian contained the scalar relativistic corrections and the spin-orbit coupling was taken into account by the
second variation method \cite{Hobbs2000}.
A Monkhorst-Pack \emph{k}-point grid of (8$\times$8$\times$1) for reciprocal space integration and 600 eV energy cut-off of the plane wave expansion have been used to get a residual error on the forces below 1 meV/\r{A} and a fully converged electronic structure including SOC.
To correctly describe the (001) surface and avoid interactions between periodic slabs, we used a supercell with a vacuum space of 12 \r{A}.  
Additionally, in order to take into account the octahedra rotations and tilting, we considered a ($\sqrt{2}\times\sqrt{2}\times$7) cell as shown in Fig. \ref{fig:STO-d}, enforcing the cubic symmetry at the cell center. 
Aiming to prevent a fake polarization along the $z$-axis, the central plane is used as $m_z$ mirror symmetry such as the slab is repeated along the $z$ direction.


\section{Results}
In what follows, the electronic characterization of the Rashba-like splitting for Sr$B$O$_3$ with $B$ = Ti and Hf, including surface oxygen vacancies at the $B$O$_2$ termination according to experiments \cite{Santander-Syro2014} is discussed. 
Great emphasis is done on the size of the splitting and the spin structure.


\subsection{STO:TiO$_2$ terminated surface:}
Fully relaxed STO surface reveals a change in the interatomic distances close to the surface due to the missing Ti--O surface  bonds (rumpling).
The Ti--Ti distance calculated between the Ti surface layer and the next just below is 3.821 \r{A}, which is much smaller than the STO bulk distance of 3.905 \r{A}. 
Additionally, the out-of-plane Ti--O bond length at the  surface changes its distance from 1.948 to 1.950 \r{A}, leading to a local polar like displacement. 
We also found a substantial lowering of the calculated electronic gap, which goes from 2.0 eV at the bulk level to 0.8 eV in the slab geometry. 
We computed the electronic band structure with and without SOC and we did not observe substantial differences, which implies that  SOC has a small contribution to the changes in the electron band structure compared with the geometrical changes.
Furthermore, no octahedral rotations or tilting were observed in the fully relaxed surface when the cubic symmetry is imposed at the central slab plane. 

 \begin{figure*}[htb]
 \centering
 \includegraphics[width=14.0cm,keepaspectratio=true]{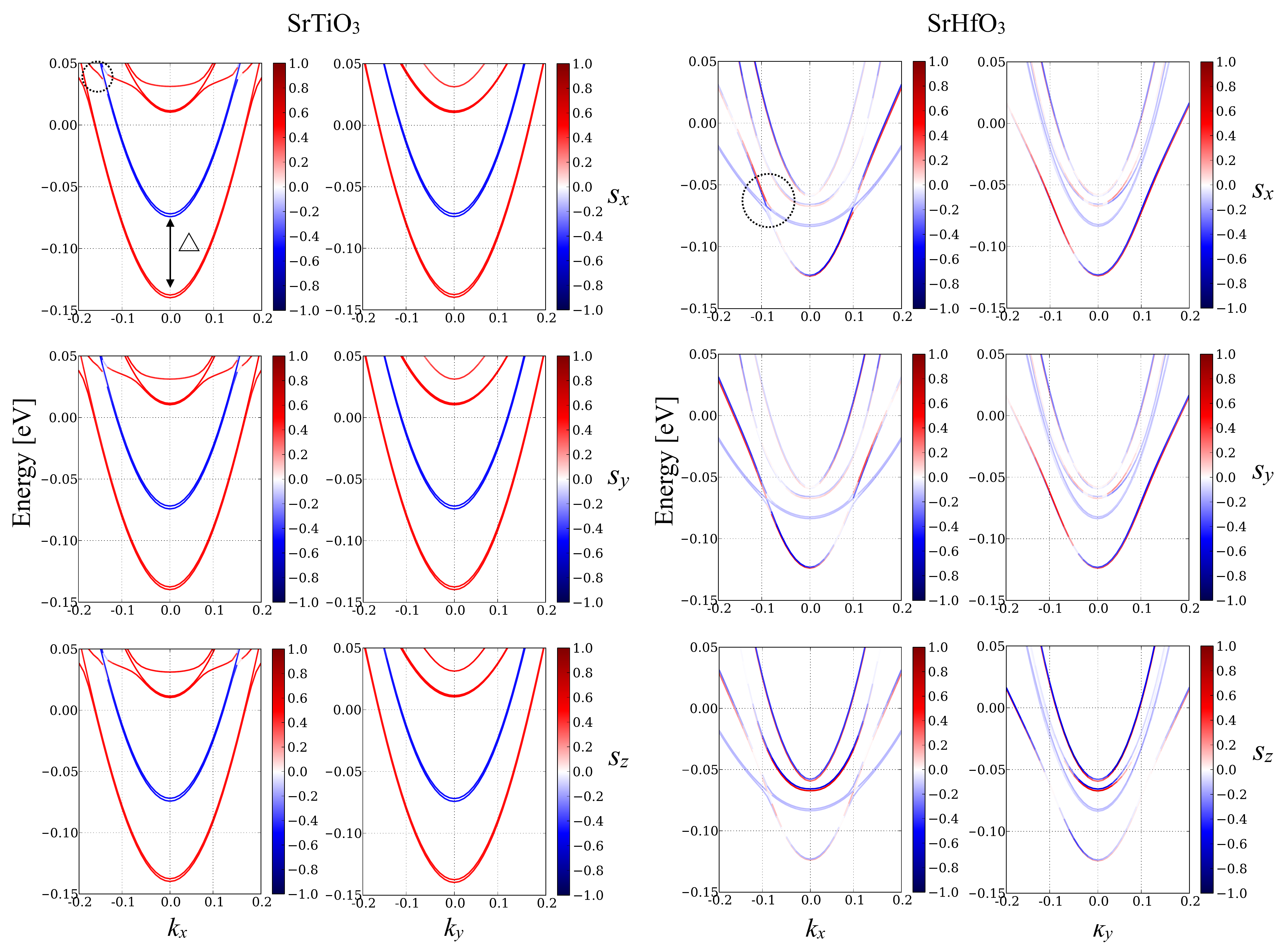}
 \caption{(Color online) Band structures around $\Gamma$-point for SrTiO$_3$ and SrHfO$_3$ with an oxygen vacancy on the $B$O$_2$ terminated surface. Red and blue denote positive and negative projection of spin vector $\vec{s}$ on Cartesian axes, units of the $k$-vector in units of $\pi/a$. The effect of the SOC and a Rashba-type splitting is observed in the \emph{d$_{xy}$} and \emph{d$_{xz/yz}$} bands crossing (noted with the dashed circles) as well as the vertical energy shift $\Delta$ $\equiv$ $\Delta E_{Zeeman}$+$\Delta E_{l}$. For SHO a spin-splitting of the lower bands is observed for the $x$ and $y$ components around E = --0.1 eV. Figures were created by using the PyProcar  script \cite{pyprocar}.}
 \label{fig:bands-spin}
\end{figure*}


\subsection{STO:TiO$_{2-\delta}$ surface with oxygen vacancies:}

Experimentally, STO surface samples become conducting in the presence of oxygen vacancies \cite{PhysRevLett.113.177601,PhysRevLett.113.086801}. 
In order to analyze their effect, we performed full structural relaxation of the STO:TiO$_{2}$ surface by removing one oxygen in the last TiO$_2$ termination layer ($\delta$ = 0.5),  imposing a bulk cubic symmetry to allow us a direct comparison with the pristine case.
As we illustrate it in Fig. \ref{fig:STO-d}, octahedral tilting and rotations are now observed in the presence of oxygen vacancies (the rotation angle is 10.9$^\circ$ with respect to the $z$-axis for the tetrahedral closest to the vacancy), in contrast to the surface without oxygen vacancies, where no rotation was observed. 

In what follows we study the electronic and spin structure of the (001) STO:TiO$_{2-\delta}$ surface. 
In Fig. \ref{fig:bands-spin}, we report the calculated  electronic band structure around the $\Gamma$--point including non-collinear magnetism and SOC. 
We found a conducting behavior very similar to what has been observed at the LaAlO$_3$/SrTiO$_3$ interface \cite{PhysRevX.3.021010}, where the lowest energy bands are the Ti--3$d$ orbitals. 
At the bulk level, the Ti--3$d$ orbitals present in the TiO$_6$ octahedra are expected to split into \emph{t$_{2\mathbf{g}}$} (\emph{d$_{xy}$}, \emph{d$_{xz}$}, \emph{d$_{yz}$}) and \emph{e$_{\mathbf{g}}$}  (\emph{d$_{z^2}$}, \emph{d$_{x^2-y^2}$}) due to the crystal-field splitting. 
Here, we remark that small distortions in the octahedral geometry, as the one observed at the surface, generate a small shifting in the energy-values of the electron levels ($\Delta E_l$) and a possible mixing of the \emph{t$_{2\mathbf{g}}$} and \emph{e$_{\mathbf{g}}$} states, mainly due to the symmetry breaking and the uncompensated TiO$_6$ octahedra at the surface \cite{PhysRevB.88.041302}. 
In the oxygen vacancy STO surface we found that the lower $d$--levels bands corresponds to the \emph{d$_{xy}$} orbitals in the paths that runs from -0.5\emph{k$_x$} -- $\Gamma$ -- +0.5\emph{k$_x$}  and -0.5\emph{k$_y$} -- $\Gamma$ -- +0.5\emph{k$_y$}. 
In addition, we also found a magnetic moment on the Ti atoms close to the surface. 
The amplitude of these magnetic moments is around 0.212 $\mu_B$/Ti-atom in the surface and 0.026 $\mu_B$/Ti-atom in the sub-surface layer. 
These results are in agreement with the  experimental observation of magnetism in STO  with oxygen vacancies \cite{Rice2014} and other theoretical works of the same system \cite{ic502905m,AFM-STO_surf,PhysRevLett.111.217601}. 
The spin polarization is analyzed by plotting the $s_x$, $s_y$ and $s_z$  components of the magnetization for each path, where up and down spin orientations are depicted by red and blue colors respectively.
We can appreciate that lower light bands close to the Fermi level are separated vertically due to an induced Zeeman splitting ($\Delta E_{Zeeman}$) as expected from the magnetism found at the surface and the energy splitting $\Delta E_l$. 
This magnetic moment is induced by the effect of the electron surface density rearrangement as a response of the uncompensated orbitals, thus oxygen vacancy acts as a magnetic impurity  \cite{PhysRevLett.111.217601,ic502905m}.

From Fig. \ref{fig:bands-spin} we can also see the parabolic-like shape of the $d$-bands, which is a characteristic of a 2DEG, as reported in previous studies \cite{Meevasana2011,Santander-Syro2011,PhysRevLett.113.086801}. 
Here, it is important to note that this 2DEG is induced by the electronic reconstruction related to the confinement of the electrons at the surface \cite{PhysRevLett.113.177601,PhysRevLett.111.217601}, rather than a polar catastrophe origin as mentioned for the SrTiO$_3$/LaAlO$_3$ interface \cite{Nick-2014}. 
Thus, we can expect that the behavior of the conducting electrons in STO differs from the one found in the oxide/oxide interfaces.

The effect of the SOC can be clearly appreciated in the Rashba-like spin-splitting at the \emph{d$_{xy}$} and \emph{d$_{xz/yz}$} crossing (pointed out by the dashed circles in Fig. \ref{fig:bands-spin}), similarly to what has been previously reported in this surface \cite{STO-quasiparticle}. 
For the \emph{d$_{xy}$} state, we obtain an effective mass of 1.6 m$_e$ for the orbital located at the second layer of TiO$_2$. 
This calculated effective mass is two times larger than the one reported in Ref.\onlinecite{STO-quasiparticle} (0.65 $m_e$), which is too small for $d$ like states \cite{doi:10.1021/cm404079a}. 
However, a spin-splitting such as in a classical Rashba picture is not observed (\emph{i.e.} crossing of the spin up and down bands) between the light branches or a spin inversion within the $d_{xy}$ bands for any of the spin-orientations along the $k_x$ and $k_y$ paths.
 This result is in contrast with the interpretation of the experimental measurements of Santander-Syro \emph{et al.}  \cite{Santander-Syro2014} who reported a giant spin-splitting in the lowest $d$-bands of the (001) STO surface.  
 One possible explanation of this splitting is that it can be a consequence of the surface ionic vibrations. 
 In that respect, we have performed ab--initio molecular dynamics calculations at constant temperature of 400 K for the slab geometry. 
Selected uncorrelated frames were taken from the simulation and electronic band structures were calculated for each case. 
 As for the 0 K ground state, we did not observe any Rashba splitting  like the one reported in Ref \onlinecite{Santander-Syro2014}, which indicates that at least for low temperatures there is no coupling between the ionic vibrations and the electronic response. 
 
 Another possibility could be that it comes from anharmonic vibrations at much higher temperatures, a possibility we do not explore in this work.
The absence of a large spin-splitting can be understood from the small SOC present in the 3$d$--Ti orbitals. 
To test this hypothesis, we performed a computer experiment by replacing all the Ti atoms by a heavier cation with much larger SOC such as Hafnium. Taking the advantage that the Hf atom has the same electronic configuration as the Ti but with 5$d$ orbitals \footnote{The spin-orbit coupling in Hf is around 11 times larger than in Ti based in the $(Ze^2/\hbar c)^2$ expression}. 
In these calculations cell lattice parameters and internal coordinates of the optimized geometry of the STO surface were fixed in order to keep the same structure. 
The 2DEG generated at the HfO$_2$ surface with oxygen vacancies presents the same main features as in the STO case. 
However, a large change in the spin-splitting state is observed in the electron dispersion as shown in Fig. \ref{fig:bands-spin}, and somehow closer to the experimental suggestion of A. F. Santander-Syro \emph{et al.} \cite{Santander-Syro2014}, who claimed for a spin-inversion direction within the lowest $d$-band.  
In  both, $k_x$ and  $k_y$ paths in Fig. \ref{fig:bands-spin} and at the Fermi level the bands are mainly composed by $s_z$ spin direction. 
The spin switching is observed for the light \emph{d$_{xy}$} Hf band at E = --0.1 eV and reciprocal points close to --0.1 and 0.1  for $s_x$  and $s_y$ spin components respectively. 
This suggests that a strong surface polarization is not enough to induce a Rashba-like spin-splitting and the presence of high-$Z$ cations in the structure is also necessary to achieve this spin state as previously reported for Ag, Cu, and Au surfaces \cite{PhysRevB.63.115415}. 
The Rashba-like splitting in the SrHfO$_3$ (SHO) is only observed for the $s_x$ and $s_y$ spin components of the $xy$-plane perpendicular to $z$-axis as expected from the Rashba Hamiltonian $\cal{H}_R=$ $\alpha_R(\vec{e}_z\times\vec{k})\cdot\vec{s}$   \cite{rashba1984}. 

\begin{figure}[htb!]
 \centering
 \includegraphics[width=9.0cm,keepaspectratio=true]{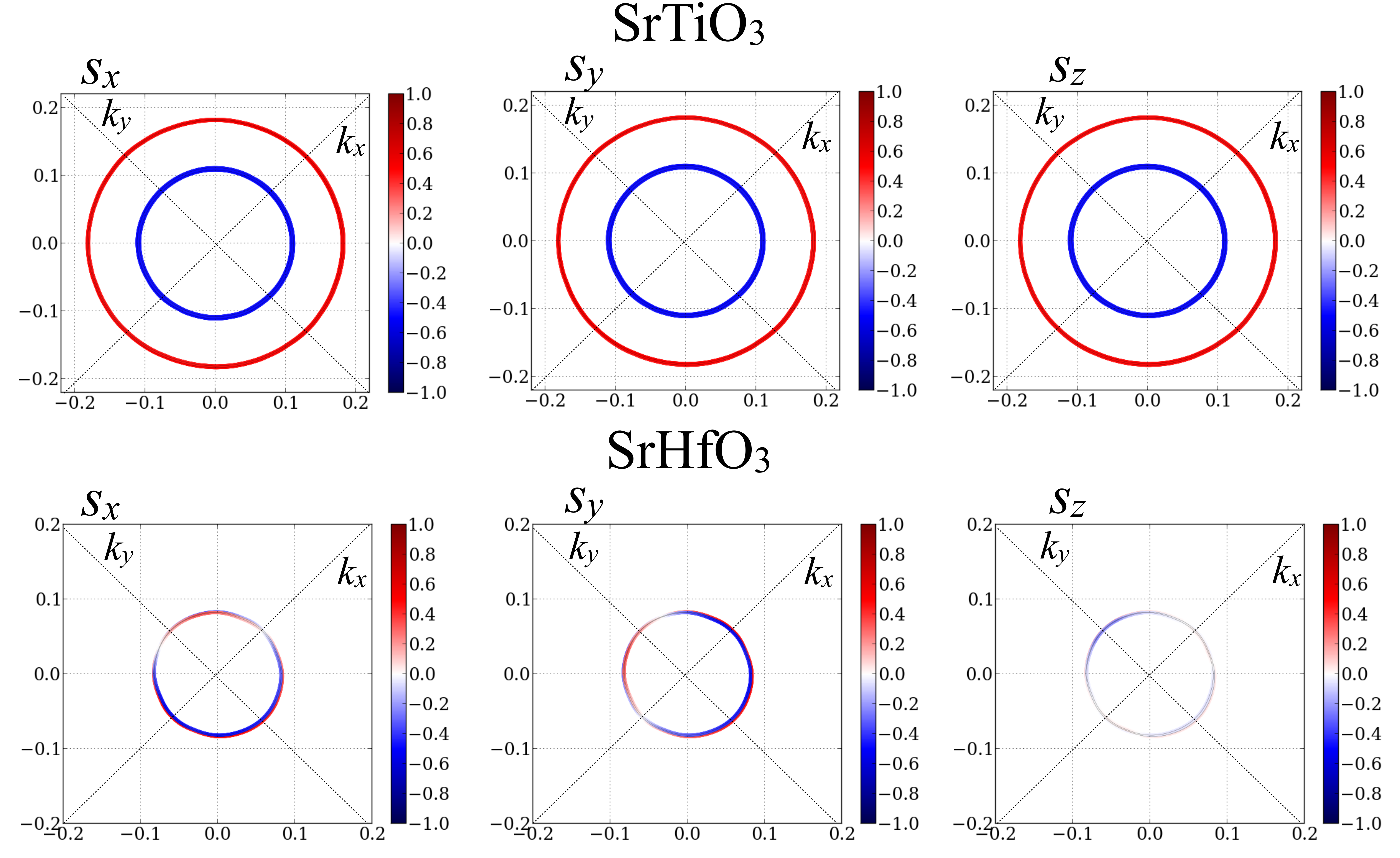}
 \caption{(Color online) Band structure at the \emph{k$_x$-k$_y$} plane (in $\pi/a$) where the spin-texture is calculated for each spin component along the cartesian axis (\emph{i.e.} $s_x$, $s_y$ and $s_z$). For SrTiO$_3$ compound (above, E = 0 eV) simple concentric rings for up (red) and down (blue) spin polarizations are observed. In contrast a Rashba-like splitting is observed by replacing the Ti cation by the heavier Hf in the SrHfO$_3$ (bottom, E = -- 0.1 eV). Figures were created by using the script PyProcar \cite{pyprocar}.}
 \label{fig:spin-texture}
\end{figure}

In addition to the band analysis, we have also extracted the in-plane spin-texture as shown in the Fig. \ref{fig:spin-texture}.
In Fig. \ref{fig:spin-texture} we can see two concentric rings for up and down spin polarization respectively for the STO (at E = 0 eV) and one single ring for SHO (E = --0.1 eV) surface. 
The rings in the STO appears as a result of the vertical splitting due to $\Delta E_{Zeeman}$+$\Delta E_l$, where $\Delta E_l$ is induced by the lowering of the \emph{d$_{xy}$} energy bands in comparison to \emph{d$_{yz}$} and \emph{d$_{zx}$}  orbitals as discussed before. 
In the case of SHO (Fig. \ref{fig:spin-texture}), spin precession in the lower ring is observed in addition to a spin switching of the \emph{d$_{xy}$} band in agreement with the bands in Fig. \ref{fig:bands-spin}) as expected for the Rashba-type splitting. 

The absence of the spin-splitting in the STO surface could be also linked to magnetism, as observed at  the  the Gd(0001) surface \cite{krupin2005} where the presence of magnetism and a Rashba-type splitting has been reported. 
For this system the splitting is lower than 0.1 \r{A}$^{-1}$ even with the large SOC present in Gd (around 9 times larger than in Ti)\cite{krupin2005}. 
Moreover, due to the presence of $\Delta E_{Zeeman}$, there is no crossing of the bands and the direction of the magnetization vector ($\vec{M}$) determine the presence of the spin-splitting. 
This is due to the fact that the Rashba-splitting vanishes for $\vec{k}\parallel \vec{M}$ and only $\Delta$E$_{Zeeman}$ remains \cite{krupin2005}. 
The later statement needs to be carefully taken into account in the STO case for which we find a magnetic moment components at the surface $m_x = $0.213, $m_y = $0.212 and $m_z = $0.212 $\mu_B$/Ti-atom respectively and thus after considering several possible magnetic orientations. 
Different starting magnetic configurations, along $x$- $y$- and $z$-axis respectively were taken into account and full structural and energy minimizations were applied. 
Nonetheless, the system always relaxed toward the same magnetic ground state in which a diagonal direction of the magnetic moment is observed with respect to the surface. 
No traces of magnetization were observed at the oxygen sites confirming the Ti--localized magnetic moments.
Thus, this rather large in- and out-of-plane magnetic moments have the tendency to prevent the possibility to have a giant Rashba splitting, in addition to the small SOC discussed previously. 

In the case of SHO, we obtain that the Hf magnetic moments are strongly reduced to $m_x=$0.034, $m_y=$0.034 and $m_z=$0.032 $\mu_B$/Hf--atom respectively close to the surface, which strongly reduced the $\Delta$E$_{Zeeman}$. 
Fig. \ref{fig:spin-texture} shows that the lowest $d$-band in SHO only exhibit spin components along the $xy$-plane. 
Nonetheless for  higher energy values, the bands are mainly spin-polarized along the $z$-axis, demonstrating a very complex spin structure with a unique spin-texture.
Therefore, only a deep understanding of this subtle competition between the polar distortions, the symmetry conditions, SOC, and magnetism would allow us to tune the Rashba-splitting at the (001) $B$O$_2$ surface. 
  
In order to test the validity of our results against the exchange-correlation functional, we also carried out the same calculations with both the Local Density Approximation \cite{LDA} (LDA) and the GGA PBEsol$+U$ functionals.
Using the LDA approximation, we did not find substantial differences with the GGA PBEsol calculations. 
Due to the strong correlation of the $d$-orbitals that drives the magnetism, we might expect the DFT+$U$ formalism \cite{lda+u} to enhance the electron correlation and change the physical picture discussed previously. 
Indeed, above $U=$ 2 eV, we observe an insulating state of the STO surface with the oxygen vacancy, which do not agree with the experimental measurements \cite{Santander-Syro2011, Santander-Syro2014,STO-quasiparticle}. 
Then, for whatever $U$ value below 2 eV, we do not observe a large Rashba+Zeeman (RZ) type spin-splitting. 
In order to test the oxygen vacancy position effects, we performed a second set of calculations considering the oxygen vacancy in the second inner layer from the surface instead of the first layer and, as before, we did not observe a Rashba spin-splitting in the STO surface within this configuration.


\subsection{\emph{Pbnm} SrHfO$_3$:HfO$_2$ surface with oxygen vacancies:} 
\begin{figure}[htb!]
 \centering
 \includegraphics[width=8.5cm,keepaspectratio=true]{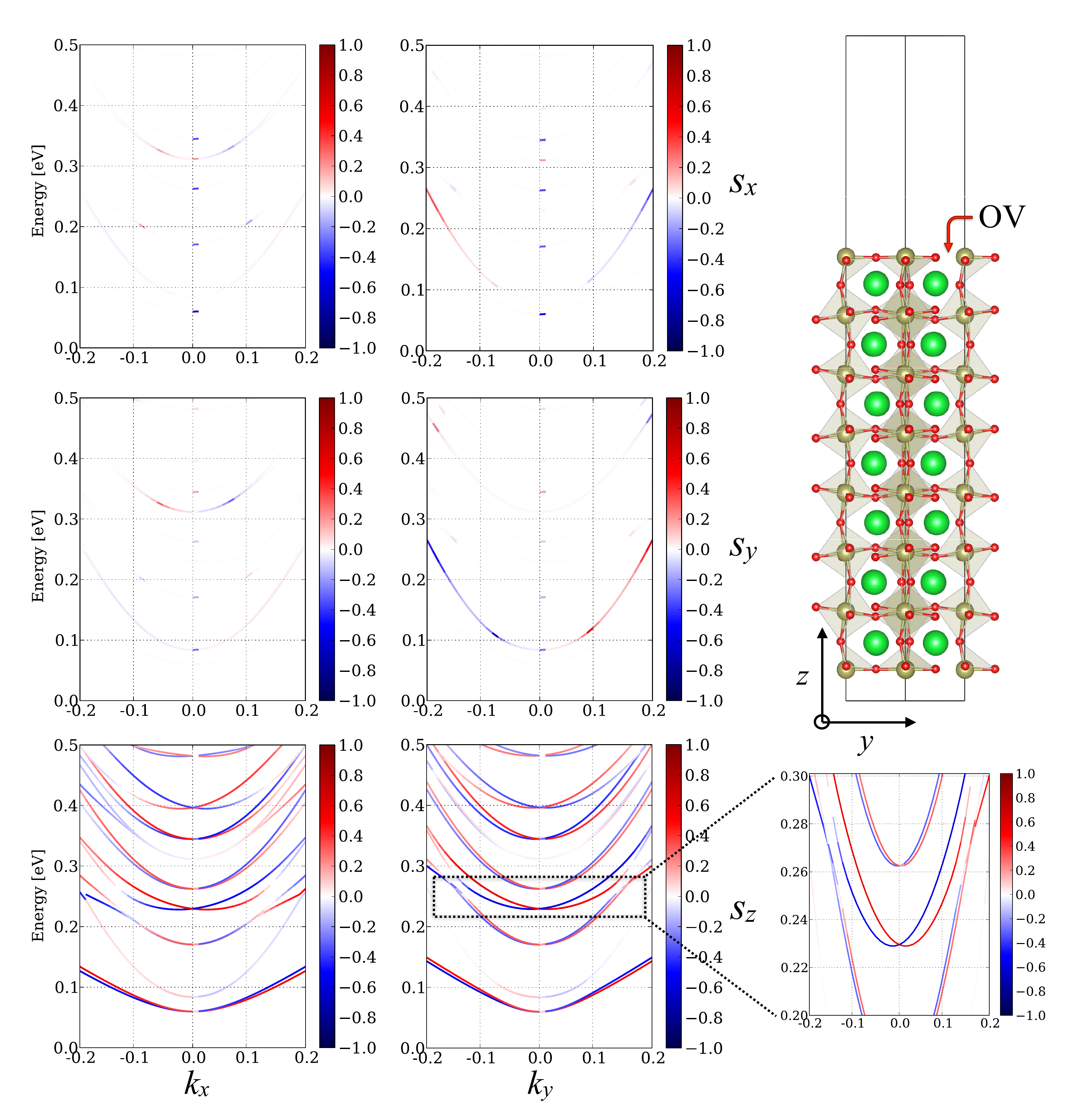}
 \caption{(Color online) Band structures around the $\Gamma$-point along $k_x$ and $k_y$ for SrHfO$_3$ surface with an oxygen vacancy on the top of the HfO$_2$ layer for each direction of the spin. Red and blue colors represent the up and down spin directions respectively. A larger Rashba-type splitting of the bands is observed for the $z$ component as appreciated for the $k_y$ -- $s_z$ extracted image. The SHO slab structure is also shown in which the large octahedral tilting can be observed. Figures were created by using the script PyProcar \cite{pyprocar}.}
 \label{fig:bands-spin-Hf}
\end{figure}

We completed the analysis of the Rashba splitting mentioned above by analysing the spin textures in the SrHfO$_3$ ground state, \textit{i.e.} in the \emph{Pbnm} phase with $a$ = 5.7516 \r{A}, $b$ = 5.7646 \r{A}, and $c$ = 8.1344 \r{A} \cite{PhysRevB.60.2972}. 
To analyze the new SHO surface in its $Pnma$ phase, we fully relaxed the surface including an oxygen vacancy in the HfO$_{2-\delta}$ termination slab. 
Our results reveal an increase of octahedral tilting and rotations as a result of the free surface plus the oxygen vacancy, which induce a structural reconstruction with respect to the bulk (see Fig \ref{fig:bands-spin-Hf}).  

From the electronic structure analysis we observe a strong reduction of the band gap, from 3.22 eV in the bulk to 0.06 eV in the surface with oxygen vacancy and thus an absence of the 2DEG.
Additionally, from the band structure plotted for each spin orientation (see Fig. \ref{fig:bands-spin-Hf}, bands around the $\Gamma$--point along $k_x$ and $k_y$ paths) we can appreciate that the lowest Hf bands corresponds to 5$d$-levels, and more precisely to the 5$d_{xy}$ states.  
According to Fig. \ref{fig:bands-spin-Hf}, we note that the major spin components for the 5$d$ bands are along the $z$-axis and a Rashba-type splitting is also observed for this orientation in most of the branches, which happen to be out-of-plane and perpendicular to the surface according to our results.
This behavior can be explained in terms of the polarization close to the surface and the Rashba Hamiltonian expressed before. 
Here, the bands present a spin-splitting for a dispersion along a $\vec{k}$-vector perpendicular to the electric field $\vec{e}_{xy}$. 
To confirm this hypothesis, we computed the spontaneous polarization by means of the Berry phase approach \cite{Vanderbilt2000147}. 
We obtained that the $P_z$ component is zero by construction and the in-plane components are equal to 0.44 $\mu$C$\cdot$cm$^{-2}$ and 1.15 $\mu$C$\cdot$cm$^{-2}$ found along the $x$-axis ($P_x$) and $y$-axis ($P_y$) respectively. 
This in-plane polarization in orthorhombic perovskites is quite well known and is related to the presence of anti-polar motion of the $A$-site in the $xy$-plane, which, in the case of the surface symmetry breaking, do not cancel out from layer to layer and thus resulting in an effective spontaneous polarization along the $xy$-plane \cite{ADFM:ADFM201300210, C5DT00010F}. 
Interestingly, and in contrast to the STO case, we do not find any presence of magnetism in the \emph{Pbnm} SHO surface with oxygen vacancies and at the same time the Rashba-splitting is considerably enhanced, confiming the rule discussed above.
Interestingly, when replacing the Hf by Ti in the relaxed $Pbnm$ SrHfO$_3$ surface, the magnetism appears again and the Rashba splitting dissapear. 
All of that confirms the influence of the magnetic ordering and the SOC strength of the $B$-site at the spin-texture of Sr$B$O$_3$ surfaces with oxygen vacancies. 


\section{Summary and Discussions}
We have analyzed from first principles calculations the (001) Sr$B$O$_3$, ($B$ = Ti and Hf) surface with oxygen vacancies. 
We found an insulator to metal transition behavior when an oxygen vacancy is included in STO at the TiO$_2$ surface, which produces an electronic charge rearrangement that results into a 2DEG located at the surface and magnetization of the Ti--$d$ orbitals close to the vacancy. 
In the limits of our DFT calculations, it is unlikely that a large Rashba-type splitting in the light lowest \emph{d$_{xy}$} bands exists in the presence of an oxygen vacancy at the STO:TiO$_2$ surfaces. 
This is in contrast with the giant spin-splitting of about 0.1 \r{A}$^{-1}$ reported by Santander-Syro \emph{et al.} \cite{Santander-Syro2014} in their experiments. 
This discrepancy can come from different origins.
On the experimental side, due to the methodology of the spin-ARPES technique used by Santander-Syro \emph{et al.}, excited states could have been induced at the surface, with a different spin-polarization of the photoelectrons. 
Then, these states can become accessible and be measured by this technique as already reported before \cite{PhysRevX.4.011046, PhysRevLett.109.097601, Jozwiak2013}. 
There is also a possibility of high temperature gradients at the surface due to the high energy from the laser field. This could create large ionic vibrations, which can create large ionic vibrations, even if the adiabatic approximation still holds \cite{PhysRevB.82.113402}.
On the calculation side, it is possible that the configuration chosen for the oxygen vacancy does not correspond to what is present experimentally. 
In that respect imaging techniques such as STM would be of use in order to check the surface structure.

It has been shown from photoemission spectroscopy that the Rashba spin-splitting in these 2DEG systems is extremely sensitive to the charge and orbital ordering, the SOC or the lattice degrees of freedom \cite{STO-quasiparticle}.
We observed such a sensitivity when replacing Ti by Hf atom with a much larger SOC, which induces a larger change in the spin-splitting landscape at fixed lattice degrees of freedom.
Indeed, when relaxing toward the SHO ground state, we observe a totally different magnetic and Rashba spin-splitting states, all of that confirming the strong spin structure sensitivity of these surfaces.
However, in all the configurations we considered, we never obtained a Rashba spin-splitting as large as the one reported by Santander-Syro \emph{et al.} and the fact that it is much larger than the one observed in the LaAlO$_3$/SrTiO$_3$ interfaces ($<$0.01 \r{A}$^{-1}$   \cite{PhysRevLett.104.126803, Hurand2015} and supported by theoretical analysis  \cite{PhysRevLett.109.196401, PhysRevB.87.161102}) call for additional studies (theoretical and experimental) in order to understand this unexpected large splitting.
Going beyond this discrepancy, we proved that sizeable Rashba-type splitting can be achieved in multifunctional perovskite oxides with large SOC, such as SrHfO$_3$ not observed before. 
These systems thus opens the possibility to tune exotic properties for new technological applications such as spin-orbitronics \cite{Kuschel2014,zhang2015}.


\begin{acknowledgments}
We acknowledge the support from the Extreme Science and Engineering Discovery Environment (XSEDE), which is supported by National Science Foundation Grant No. ACI-1053575. AHR acknowledge the support of DMREF-NSF 1434897 and the Donors of the American Chemical Society Petroleum Research Fund for partial support of this research under contract 54075-ND10. EB thanks the FRS-FNRS and the Consortium des Equipements de Calcul Intensif (CECI) funded by the FRS-FNRS (grant 2.5020.11) and the PRACE project TheDeNoMo. MGV acknowledge partial support from the Basque Country Government, Departamento de Educaci\'on, Universidades e Investigaci\'on (Grant No. IT-756-13), the Spanish Ministerio de Econom\'ia e Innovaci\'on (Grant No. FIS2010-19609-C02-01 and FIS2013-48286-C2-1-P) and the FEDER funding.
\end{acknowledgments}

\bibliography{./paper}

\end{document}